\documentclass[referee,usenatbib]{mn2e} 
\usepackage{amssymb}
\usepackage{subfigure}
\usepackage{multirow}
\usepackage{graphicx}
\title[The {\it{XMM-Newton}} view of RLNLS1 galaxy PMN~J0948+0022]{The {\it {XMM-Newton}} view of the radio loud narrow line Seyfert1 galaxy PMN~J0948+0022}
\author[Subir Bhattacharyya, Himali Bhatt, Nilay Bhatt and Krishna Kumar Singh]{Subir Bhattacharyya\thanks{E-mail:subirb@barc.gov.in}, Himali Bhatt, Nilay Bhatt and Krishna Kumar Singh \\
Astrophysical Sciences Division, Bhabha Atomic Research Centre, Mumbai 400085, India}
\begin{document}
\newcommand {\pmn}{{PMN~J0948+0022}~}
\newcommand {\fermi}{{\it{Fermi}}~}
\newcommand {\xmm}{{\it{XMM-Newton}}~}
\newcommand {\apj}{ApJ}
\newcommand {\apjs}{ApJS}
\newcommand {\apjl}{ApJL}
\newcommand {\aap}{A\&A}
\newcommand {\mnras}{MNRAS}
\newcommand {\nat}{Nature}
\newcommand {\aaps}{A\&AS}
\newcommand {\pasj}{PASJ}
\date{Accepted 1988 December 15. Received 1988 December 14; in original form 1988 October 11}

\pagerange{\pageref{firstpage}--\pageref{lastpage}} \pubyear{2002}

\maketitle

\label{firstpage}

\begin{abstract}
We analysed the archival \xmm data of the radio loud narrow line
Seyfert1 galaxy  PMN~J0948+0022 in the energy range 0.3--10.0 keV.
The X-ray data reveal that the spectrum in the 0.3--10.0 keV energy band is not a simple power-law as previously 
described in the literature. Instead it consists of a power-law with soft excess below 2.5 keV. The X-ray spectrum was
fitted with four different models and it was shown the soft excess component of the spectrum in the 0.3--2.5 keV
energy range could be described reasonably well within the framework of the thermal Comptonization model as well
as relativistically blurred reflection model. The power-law component required to fit the spectrum beyond 2.5 keV
was found to be rather hard compared to the ones observed in other unobscured active galactic nuclei. It is also shown
that the {\it Swift}/XRT spectrum from the source could not reveal the soft excess component due to poor statistics.
The fractional variability estimated from \xmm data indicates the presence of independently varying 
components in the spectrum above and below 1 keV.
\end{abstract}

\begin{keywords}
extragalactic : Seyfert1 -- X-ray -- individual:PMN J0948+0022 
\end{keywords}

\section{Introduction}
\pmn is a radio-loud narrow line Seyfert~1 (RLNLS1) galaxy, with cosmological redshift $z\,=\,0.5846$ and radio loudness 
parameter $R \approx 846$, defined as the ratio of the flux at 1.4 GHz to the flux at 440 nm (Radio loudness criterion : R$>$10) 
\citep{fos11},  detected in $\gamma$-rays by Large Area Telescope (LAT) on board \fermi satellite \citep{abdo+09b, abdo+09a}. 
 It is 
also one of the six such sources detected by \fermi/LAT in the MeV--GeV 
region\footnote{http://www.asdc.asi.it/fermi2lac/}. \citet{abdo+09b, abdo+09a}
reported the multi-wavelength observations of \pmn during March--July 2009, while 
\citet{fos+11} observed the source during a $\gamma$-ray outburst in July 2010. In
both cases, the observed spectral energy distribution (SED)  looks very similar to the
SED of typical high power blazars, known as flat spectrum radio quasars (FSRQs). The SEDs were modeled with the synchrotron
and Comptonization models developed by \citet{gt09}. For both the
observations, the broadband SED were fitted with a synchrotron component for 
radio--to--IR/optical, a synchrotron self-Compton component for the X-rays, external 
Comptonization component for MeV--GeV $\gamma$-rays, and a thermal component from the
accretion disk in the UV band. Interestingly, the fitted parameters and estimated power
were found to fall in the region between the low power blazars (BL Lacs) and high 
power blazars (FSRQs). These derived properties are similar to those of other three \fermi 
observed RLNLS1 \citep{abdo+09b}. Their analysis of the broadband SED 
lead to similar conclusions. But the most intriguing fact is that the estimated 
black hole (BH) mass and mass accretion rates for RLNLS1, in terms of Eddington rate, are very different 
compared to those properties of blazars. The BH mass for RLNLS1 is approximately 
two orders of magnitude lower than that of blazars (Foschini 2011). Secondly, the RLNLS1s 
accrete at near Eddington limit, while blazars accrete at much lower rate.

\noindent Here in this work, we gave a closer look at the X-ray spectrum of \pmn. The 
reason is the following. In the broadband SED modelling for \pmn ,
the X-ray component is described by synchrotron self-Compton emission
process \citep{abdo+09b, abdo+09a}. The X-ray spectra of radio quiet narrow line 
Seyfert~1 galaxies, as observed with {\xmm}, show the presence of soft excess in  
0.3--2.0 keV energy band and a power-law component above 2.0 keV \citep{bz04, bol+03, fab+04, ha+08, bri+01, pg+03, dew+07, gal+13, fab+13}. 
The origin of the soft excess 
is not properly understood yet. In fact, some radio emitting NLS1 which are not formally categorised as radio-loud, also show strong soft excess
in their X-ray spectra, e.g, PKS~0558--504 \citep{pa+10, wa+13}. 
PKS~2004--447, another radio-loud (R = 6358) NLS1 \citep{fos11}, also shows a strong soft excess in its X-ray spectrum \citep{ga+06}, 
although there is some doubt over the 
classification of this source \citep{os+01}. If the X-ray spectrum of the $\gamma$-ray 
loud RLNLS1 \pmn also shows 
the presence of a soft excess, then it is important to understand its origin. 
Here, we present an analysis 
and modelling of archival data of \pmn observed with \xmm, {\it Swift}/XRT to
show that the X-ray spectrum of the source has a complex structure, which the current 
{\it Swift}/XRT spectrum can not adequately describe. We also analysed the {\fermi}/LAT
data taken over six months around the \xmm observation to study the $\gamma$-ray emission 
from \pmn. 

\noindent The paper is structured as follows. In section 2, we described the data 
reduction procedure for \xmm, {\it Swift}/XRT and \fermi. The spectral modelling is 
described in section 3. Results are discussed and concluded in sections 4 and 5, 
respectively.
 
\section{Data Reduction}
\label{sec:redu}

\begin{table}
\centering
\begin{tabular}{ll}
\hline                  
\hline
Mission                        &  XMM-Newton\\
\hline
Observation ID                 &  0502061001 (Obs-A)  \\
Start Time (UT)                &  2008-04-29 15:12:02 (PN)\\
Exposure Time (ks)             &  24.27 \\                 
Sum of good time intervals (ks)&  12.83 (PN) \\
EPIC filter                    &  Medium\\
Mode                           &  PrimeLargeWindow (PN) \\      
\hline
Observation ID                 &  0673730101 (Obs-B) \\
Start Time for PN  (UT)        &  2011-05-28 11:50:06\\
Start Time for MOS (UT) -- M1  &  2011-05-28 11:21:36\\
Start Time for MOS (UT) -- M2  &  2011-05-28 16:20:22\\
Start Time for MOS (UT) -- M3  &  2011-05-29 00:59:12 \\
Start Time for MOS (UT) -- M4  &  2011-05-29 08:34:34\\
Start Time for MOS (UT) -- M5  &  2011-05-29 10:16:19 \\ 
Exposure Time (ks)             &  92.92 (PN); 15.82 (M1), 28.42 (M2), 24.85 (M3), 4.02 (M4), 10.14 (M5)\\                 
Sum of good time intervals (ks)&  88.00 (PN); 15.27 (M1), 28.32 (M2), 24.70 (M3), 4.01 (M4), 10.11 (M5)\\     
EPIC filter                    &  Medium\\
Mode                           &  PrimeLargeWindow (PN) ; PrimePartialW3 (MOS)\\       
\hline
Mission                        &  Swift -- XRT\\
\hline
Observation ID                 &  00031306014 \\ 
Start Time (UT)                &  2011-05-28 01:56:57\\
Exposure Time (ks)             &  3.64  \\                
Mode                           &  Photon Counting  \\     
\hline
Mission                        &  Fermi - LAT\\
\hline
Start Time (UT)                &  2011-02-27 23:59:58\\
Exposure Time (ks)             &  15638.4   \\
\hline 
\end{tabular}
\caption{Details of Observations}
\label{tab1}
\end{table}

\subsection{\xmm observations}
\pmn was observed by \xmm telescope \citep{st+01, tu+01} 
operated with Medium filter at two epochs 
during April, 2008 (Observation ID 0502061001, hereafter Obs-A) and May, 2011 (Observation ID 0673730101, hereafter Obs-B) for
approximately 24 ks and 93 ks, respectively. The X-ray data were taken in prime partial W3
mode for EPIC-MOS detectors (MOS1 and MOS2; \citet{tu+01}) for Obs-A and the source did
not fall in the active field of view, therefore,
only EPIC-PN (European Photo Imaging Camera; \citet{st+01}) data were used for further analysis.
For Obs-B, the observation with MOS detectors were taken in five short duration 
segments of 16 ks (M1), 28 ks (M2), 25 ks (M3), 4 ks (M4) and 10 ks (M5), however, 
the data were taken for full observation time span, i.e. 93 ks, for PN detector.

\subsubsection{EPIC data}
\label{sec:epic}

Data reduction followed standard procedures using \xmm Science Analysis System
software (SAS version SAS11.0.0) with updated calibration files.
Event files for the MOS \emph{(for Obs-B)} and the PN 
\emph{(for Obs-A and Obs-B)} detectors were generated  using the tasks
{\sc Emchain} and {\sc Epchain}, respectively. These tasks allow calibration of the
energy and the astrometry of the events registered in each CCD chip and to
combine them in a single data file.
Event list files were extracted using the SAS task {\sc Evselect}.
Data from the three cameras were individually screened for the time intervals
with high background when the total count rate (for single
events of energy above 10 keV) in the instruments exceed 0.35 and 1.0
$\rm{counts~s^{-1}}$ for the MOS and PN detectors, respectively.
\noindent Source photons were extracted from a circular region of radius 40$^{\prime\prime}$ around the source position
to generate the light curves and spectra. Background was estimated from four source free regions with radius of 20$^{\prime\prime}$
each located on the same CCD near the source.
The X-ray spectra and light curves of the source were generated using SAS task {\sc Evselect} and
background scaling factors have been calculated using {\sc Backscale} task.
The photon redistribution as well as the ancillary matrices were computed using SAS task  {\sc Rmfgen} and  {\sc Arfgen}.
Finally, the PN spectra of both Obs-A and Obs-B were rebinned to have at least 150 counts per bin.
In the case of ObsB, the MOS1 and MOS2 spectra from all time segments were added to generate a single spectrum using the {\sc Addspec} utility.
The final MOS spectrum was rebinned to have at least 130 counts per spectral bin. 

\subsection{{\it Swift} observations}
X-ray data from X-ray telescope (XRT) on {\it Swift} \citep{ge+04} have also been analysed here in photon counting mode.
 These observations have been taken on the same date as for
Observation ID 0673730101 from \xmm, i.e., 28 May 2011.
The {\it Swift}/XRT data have been reduced using the HEAsoft packages (6.12 version), 
including {\sc Xselect}, {\sc Ximage}, and {\it Swift} data analysis tools.
The spectrum was generated using the {\sc XRT} tool {\sc Xrtpipeline} and the  
response matrices files (RMF) and Ancillary Response Files (ARFs) were generated using task {\sc Xrtmkarf} and the {\sc Caldb} 
version 014 (released date 25 July 2011). Finally, the spectra were grouped using the {\sc Grppha} task with 30 counts per bin. 

\begin{figure*}
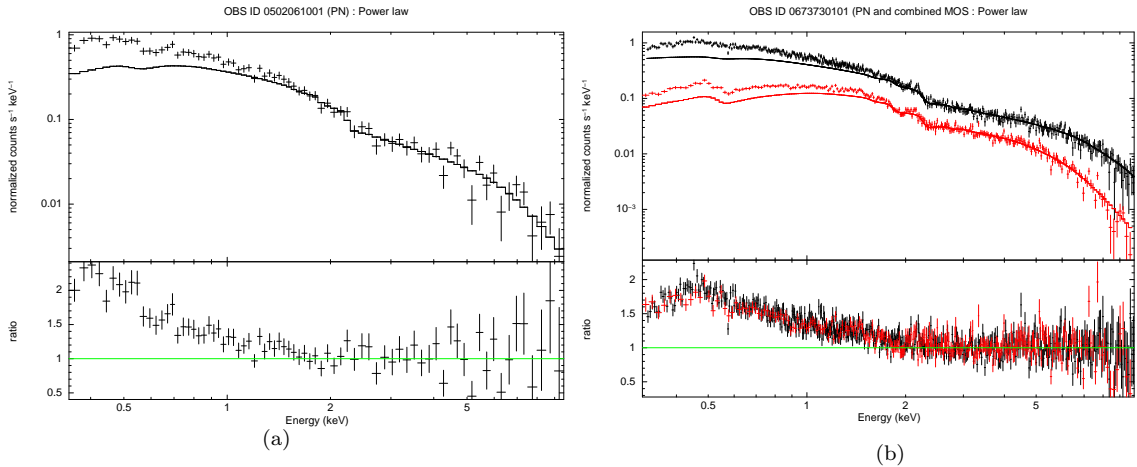

\begin{center}
  \subfigure[]{
    \includegraphics[height=0.31\textheight,angle=-90]{./ObsA_pn_fit_final_ratio.ps}} 
  \subfigure[]{
    \includegraphics[height=0.35\textheight,angle=-90]{./ObsB_pn_mos.ps} }
    \caption{Soft excess for (a) EPIC-PN spectrum for Obs-A, (b) EPIC-PN (\emph{black}) and combined MOS (\emph{red}) spectra for 
             Obs-B. The power-law model is fitted for the energy range 2.5--10.0 keV and extrapolated to lower energies to highlight the presence of soft excess in each case.} \label{fig1}
\end{center}
\end{figure*}

\subsection{\fermi/LAT observation} 
The Large Area Telescope (LAT), onboard \fermi gamma-ray observatory, is a pair 
conversion detector \citep{at+09}.
It is sensitive to the gamma-rays in the energy range $\sim$20 MeV to $>$300 GeV with peak effective
area of $\sim$8000 cm$^2$ at 1 GeV. LAT has a very wide
field of view of 2.4 sr covering 20\% of the sky at any instant with angular resolution of $\sim$1 deg. 
at 1 GeV.  
\begin{figure*}
\begin{center}
  \centering
  \subfigure[]{
    \includegraphics[width=0.4\textwidth,angle=-90]{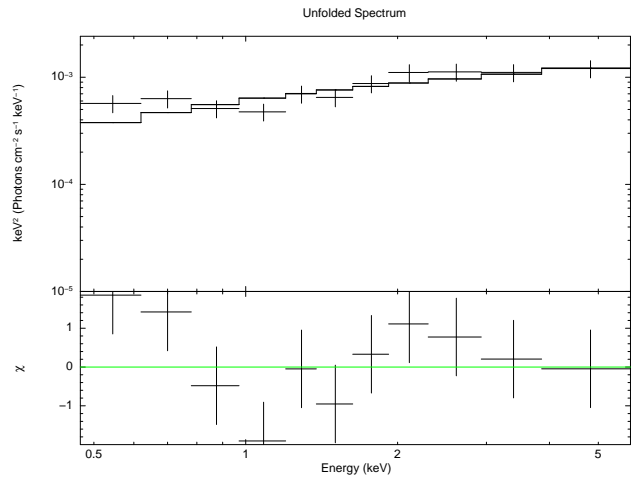}} 
  \subfigure[]{
    \includegraphics[width=0.4\textwidth,angle=-90]{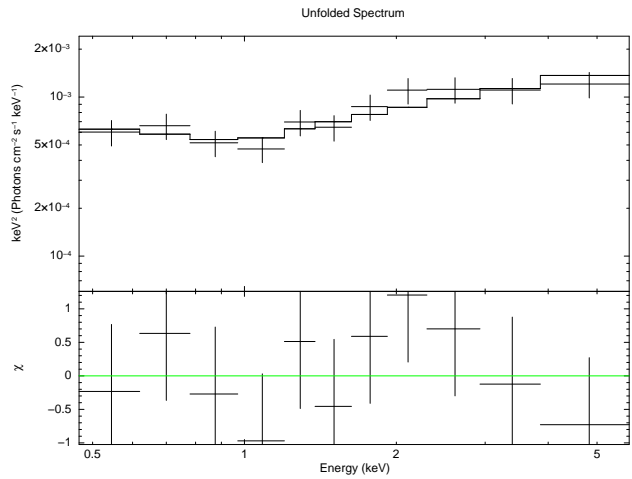}} 
  \subfigure[]{
    \includegraphics[width=0.4\textwidth,angle=-90]{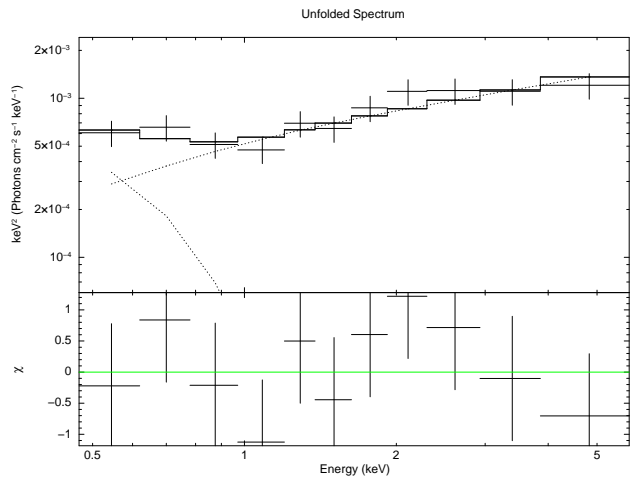}}
    \caption{{\it Swift}/XRT spectrum with fitted with three phenomenological models, (a) power-law, (b) broken power-law and (c) blackbody and power-law. It is evident that the data are consistent with the presence of the soft excess seen by \xmm, but were not 
sufficient to provide a statistical detection.}\label{fig2}
\end{center}
\end{figure*}

\noindent The LAT  data for the source considered for the present analysis cover the period 
February 28--August 28, 2011. The data have been analyzed  using the standard \fermi-LAT ScienceTools 
software package (ver. v9r27p1){\footnote{http://fermi.gsfc.nasa.gov/ssc/data/analysis/software/}} and
with the help of the analysis threads and other documentation available from the \fermi Science Support Center
webpages{\footnote{http://fermi.gsfc.nasa.gov/ssc/data/analysis/}}. 
We have selected only "Diffuse" class events (i.e. the purest gamma-ray
events) in the energy range 100 MeV to 300 GeV within 10$^{\circ}$ region of interest (ROI) 
centered around the source position 
(RA = $09^h48^m57^s.3$ and Dec = $+00^{\circ}22^{\prime}25^{\prime\prime}.6$) using {\sc gtselect} task.
In order to avoid background contamination from the bright Earth limb, we have discarded
photons arriving from zenith angle greater than $105^{\circ}$. We selected  only those time intervals
during which the events/photons qualify for science analysis using the filter expression 
(DATA\_QUAL==1 \&\& LAT\_CONFIG==1 \&\& ABS(ROCK\_ANGLE)$<$52) within {\sc gtmktime} task. 
Then the livetime for the source
was calculated and exposure map of the source was generated. Finally, using Unbinned 
likelihood analysis tool ({\sc gtlike}), we fitted the source spectrum along with other 21 sources
present within the ROI of 10$^o$ centered around PMN J0948+0022 and obtained the flux in the energy
band 100 MeV -- 300 GeV. For this, we created source model file containing model spectra for all 
these 21 gamma-ray sources visible in the ROI as listed in the 2nd \fermi/LAT catalogue and 
diffuse emission models of galactic and extragalactic origins. 
For PMN J0948+0022, we have taken power-law spectrum model.
We obtained the value of power-law index as  2.457 $\pm$ 0.001 with 
test statistics (TS; detection significance of the source $\approx \sqrt{TS}$) of 399  for the whole energy range 100 MeV--300 GeV.
The photon flux has been estimated to be 136.32$\pm$0.22 $\times 10^{-9}$ $\rm{photons~cm^{-2}~s^{-1}}$ in energy range  100 MeV--300 GeV.
The fluxes are also estimated in different energy bands and given in Table~\ref{tab:fermi}. 

\begin{table}
\centering
\begin{tabular}{lll}
\hline
Energy band                    &  Photon Flux                                    &      Energy Flux         \\
 (MeV)                         & $10^{-8}$$\rm{photons~cm^{-2}~s^{-1}}$   &    $10^{-13}$$\rm{erg~cm^{-2}~s^{-1}}$                       \\
\hline 
100--400                       & 9.15 $\pm$ 0.27                                 &    252.5$\pm$ 7.5 \\
400--1500                      & 1.92 $\pm$ 0.17                                 &    207.9$\pm$18.5 \\
1500--5500                     & 0.62 $\pm$ 0.04                                 &     65.3$\pm$15.3  \\
5500--300000                   &$<$7.70  $\times 10^{-6}$                        &    $<$0.02 \\   
\hline 
\end{tabular}
\caption{Six months averaged fluxes in different energy bands for \emph{Fermi}/LAT observations of \pmn.}
\label{tab:fermi}
\end{table}

\section{Results}
The EPIC-PN spectra of both the observations (Obs-A and Obs-B as listed in Table \ref{tab1}) 
and the combined MOS spectrum of Obs-B were fitted with a power-law 
distribution with the photon index $\Gamma$ using the spectral modelling tool {\sc Xspec} \citep{ar96}. 
The absorption model {\sc wabs} \citep{MM83} is used to account for the Galactic 
photo-electric absorption of X-rays.
The value of equivalent hydrogen column density parameter was set to $4.4\times10^{20}$ cm$^{-2}$ 
as given by the nH column density calculator \citep{ka+05} available
on HEASARC website\footnote{http://heasarc.gsfc.nasa.gov/cgi-bin/Tools/w3nh/w3nh.pl}.
This value is kept frozen for the subsequent spectral analysis presented here.
To improve the statistics, the PN and the MOS spectra of Obs-B
were fitted simultaneously. The relative cross-calibration of PN and MOS detectors was taken 
care of by introducing a floating normalization constant in the model during the fitting process. 
This value is never more than 5\% of unity.  The complete photon spectra for both the observations in the 
0.30--10.0 keV energy range could not be fitted with the power-law and galactic absorption. On the
other hand, the spectra were fitted well by a power-law in the energy range 2.5--10.0 keV. 
The photon indices were found to be 1.49$\pm$0.16 for PN spectrum in Obs-A and 1.38$\pm$0.02 
for PN and MOS spectra in Obs-B.
The $\chi^2$/dof values are 31/24 (PN, Obs-A) and 365/314 (PN and MOS, Obs-B). 
The power-law component  when extended to lower energies (down to 0.3 keV)
revealed a soft excess in the 0.3--2.5 keV energy range. The presence of soft excess component in the PN and MOS spectra are 
shown in Figure \ref{fig1}. 

\noindent 
The origin of the soft excess in X-rays observed in case of narrow line Seyfert~1 galaxies 
is still not well understood.
Here, we use different phenomenological and physical models to
explain the observed soft excess for this source.
\begin{table}
\centering
\begin{tabular}{llll}
\hline 
\hline
\multicolumn{1}{c}{Parameters}              & \multicolumn{2}{c}{{\it Xmm-Newton- (PN+MOS)}}                                                     &       \multicolumn{1}{c}{Swift/XRT} \\        
                                            &  \multicolumn{1}{c}{OBS A}                   &    \multicolumn{1}{c}{OBS B}                        &                           \\
\hline
                      Model                 & \multicolumn{1}{l}{wabs*bknpower}            & \multicolumn{1}{l}{constant*wabs*bknpower}          &  \multicolumn{1}{l}{wabs*bknpower}  \\
                 ${\Gamma}_1$               & $2.2_{-0.06}^{+0.07}$                        &  2.17$\pm$0.02                                      &  2.76$^{+0.62}_{-0.44}$                      \\ 
                  $E_{break}$ (keV)         & $1.14\pm0.11$                                &  $1.28_{-0.04}^{+0.05}$                             &  0.96$\pm$0.16                      \\
                 ${\Gamma}_2$               & $1.61\pm0.05$                                &  $1.56_{-0.02}^{+0.01}$                             &  1.45$\pm$0.12                      \\ 
                 $N_{pl}$                   & $7.68_{-0.27}^{+0.21}\times10^{-4}$          &  $5.79\pm0.05~\times10^{-4}$                        &  $5.56_{-1.53}^{+0.91}\times10^{-4}$\\ 
                     CF                     &                                              &  1.040$\pm$0.006                                    &                                      \\
               $\chi^2 / \nu$               & 80/73                                        &  1045.64/768                                        &  4.78/7                                \\
                  Log(Flux) [0.3-10.0 keV]  &  -11.226$\pm$0.005                           &  -11.354$\pm$0.001                                  &  -11.215$\pm$0.024                    \\ 
 \hline 
                      Model                 & \multicolumn{1}{l}{wabs*(zbbody+zpowerlw)}   &  \multicolumn{1}{l}{constant*wabs*(zbbody+zpowerlw)}&   \multicolumn{1}{l}{wabs*(zbbody+zpowerlw)} \\ 
                      kT (keV)              & 0.15$\pm$0.01                                &  0.173$\pm$0.003                                     &   0.14$\pm$0.04                              \\ 
                    $N_{bbody}$             & $2.62_{-0.30}^{+0.38}\times10^{-5}$          &  $1.99_{-0.06}^{+0.04}\times10^{-5}$                &   $4.89_{-2.14}^{+10.09}\times10^{-5}$       \\
                     $\Gamma$               & $1.64_{-0.04}^{+0.03}$                       &  $1.585_{-0.005}^{+0.007}$                           &   1.44$\pm$0.13                              \\   
                    $N_{pl}$                & $1.56\pm0.06~\times10^{-3}$                  &  $1.06\pm0.01~\times10^{-3}$                        &  $1.12_{-0.18}^{+0.20}\times10^{-3}$          \\ 
                     CF                     &                                              &  1.040$\pm$0.006                                    &                                              \\
               $\chi^2 / \nu$               & 79.6/73                                      &  1033.68/768                                        &   5.38/7                                        \\
                  Log(Flux) [0.3-10.0 keV]  &  -11.233$\pm$0.006                           &   -11.360$\pm$0.001                                 &   -11.223$\pm$0.025                           \\
\hline
	\end{tabular}
	\caption{The best-fit values of the spectral parameters for the two phenomenological models ({\it i}) broken power-law and ({\it ii}) (black-body + powerlaw) derived from the spectral fitting of \emph{XMM-Newton} and \emph{Swift}/XRT data.}
	\label{tab2}
\end{table}

\subsection{Phenomenological models}
\noindent Before using the physical models to fit the spectra, we used phenomenological models
-- (\emph{i}) broken power-law and (\emph{ii}) blackbody with power-law, to understand the basic features of the spectra.
The broken power-law model gives an idea about the relative steepness of the soft and the hard components of the spectrum. 
The fitting of the PN spectrum of Obs-A with a broken power-law gave
the break energy at 1.14$\pm$0.11 keV with photon indices 2.21$\pm$0.06 and 1.61$\pm$0.05 below and above the break energy, respectively. The 
$\chi^2$/dof came out to be 80.23/73. For the PN and the combined MOS spectra of Obs-B, the photon indices were found to be 2.17$\pm$0.02 and 
1.56$\pm$0.02 below and above the break energy at 1.28$\pm$0.04 keV. The $\chi^2$/dof was 1045/768. Thus the spectral features 
during the two observations were very similar. 

\noindent In the other phenomelogical model, a blackbody is used to fit the soft excess component and a power-law is
used to fit the hard component of the spectrum. The underlying assumption is that the soft excess has thermal origin. The complete model was {\sc constant*wabs*(zbbody+zpowerlaw)}. 
The fitting yielded a blackbody temperature of 0.152$\pm$0.013 keV and photon spectral index of 1.64$\pm$0.03 for the PN spectrum
of Obs-A. For the PN and the MOS spectra of Obs-B, the fitted parameters were 0.173$\pm$0.004 keV and 1.585$\pm$0.004 for blackbody temperature and 
photon power-law index respectively. The $\chi^2/dof$ were 79.61/73 and 1033.66/768 for Obs-A and Obs-B, respectively. The details of
the fitted parameters are given in Table \ref{tab2}. The values of parameters obtained in the above phenomenological models are consistent with 
the values reported for other AGN in the literature \citep{gd04, por+04, min+09}.

\noindent We fitted the {\it Swift}/XRT data with 
a simple power-law in the 0.3--10.0 keV energy range. The model used was {\sc wabs*zpowerlaw}. The fitted photon index was $1.6\,\pm\,0.1$ and 
$\chi^2$/dof=12.21/9. The equivalent hydrogen column 
density was kept fixed at $4.4\,\times\,10^{20}$ cm$^2$. Even though the soft excess could not be convincingly detected in the spectrum due to the
poor statistics of the data, the fit improves when we use more complex phenomenological models, such as,  broken power-law ({\sc wabs*bknpow}) or 
a black body spectrum along with a power-law ({\sc wabs*(zbbody+pow)}). In case of broken power-law the photon indices came as 2.76$^{+0.62}_{-0.44}$
and 1.45$\pm$0.12 below and above the break energy at 0.96$\pm$0.16 keV, respectively with $\chi^2/dof$ = 4.78/11. Similarly for the second phenomenological
model, the blackbody photon temperature came as 0.14$\pm$0.04 keV and the photon index came as 1.44$\pm$0.13. The $\chi^2/dof$ was 5.38/11. 
The fitted spectra are shown in Fig \ref{fig2}. The parameters of
the phenomenological models are consistent with that obtained with the {\sc xmm-Newton} data. 
The details of the fitting parameters are given in Table \ref{tab2}.
In the following, we limit our further analysis using more physical models to Obs-B, owing to the higher signal-to-noise ratio of these data.
\begin{figure*}
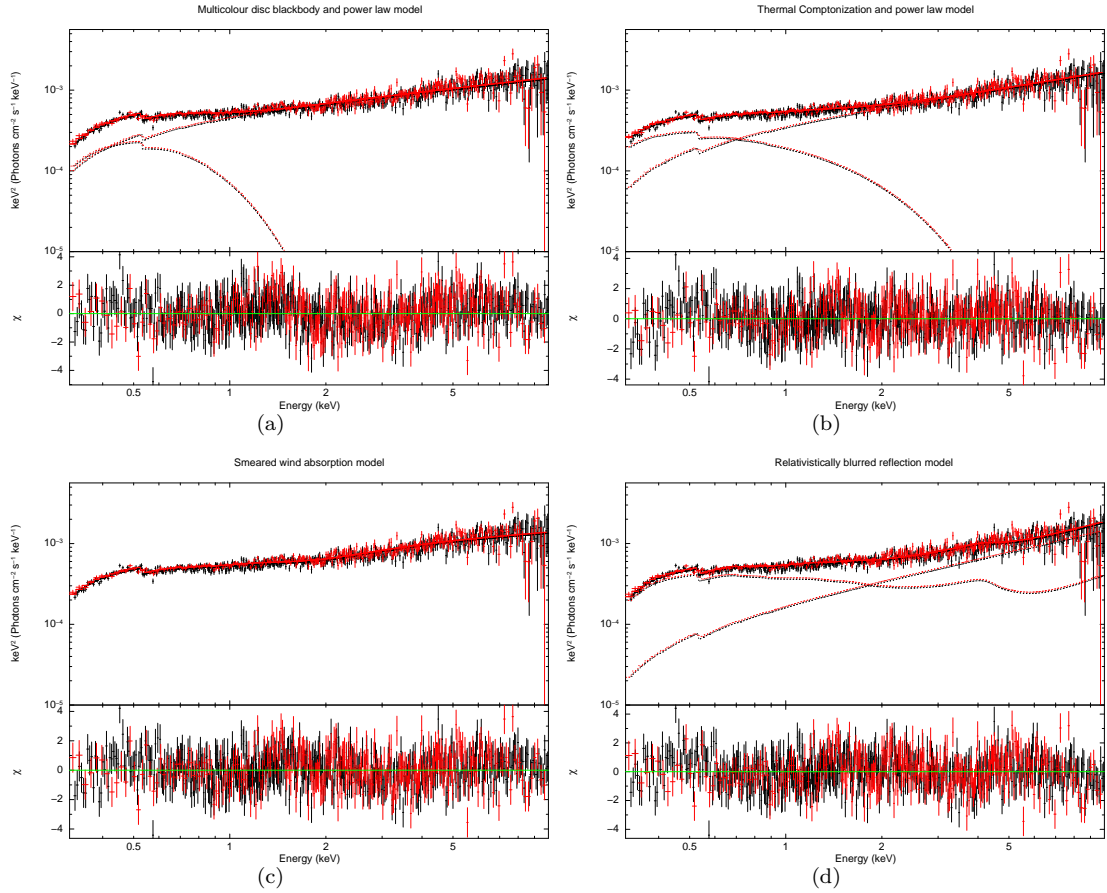

\begin{center}
  \subfigure[]{
    \includegraphics[height=0.3\textheight,angle=-90]{./spec-diskbb_pn_mos.ps}} 
  \subfigure[]{
    \includegraphics[height=0.3\textheight,angle=-90]{./spec-comptt_disk.ps}} 
  \subfigure[]{
    \includegraphics[height=0.3\textheight,angle=-90]{./specfit-swind_pn-mos.ps}} 
  \subfigure[]{
    \includegraphics[height=0.3\textheight,angle=-90]{./specfit-reflionx_pn-mos_latest.ps}} 
    \caption{EPIC- PN (\emph{black}) and combined MOS (\emph{red}) spectra of Obs-B  fitted with different  
models : (a) Multicolour disc blackbody with power-law, (b) Thermal Comptonization and power-law (disc geometry), 
(c) Smeared wind absorption model and (d) Relativistically blurred reflection model and power-law. 
$\chi^2$ distributions of fitting are given in lower panels of each plot.} \label{fig3}
\end{center}
\end{figure*}

\subsection{Physical models} 
\subsubsection{Multicolour disc blackbody (MCD) and power-law (PL)}
\noindent Considering the possibility that the soft excess is contributed by the optically thick and geometrically thin 
accretion disc of the galaxy around the central black hole,
we fitted the soft excess with a multicolour disc blackbody, using the {\sc diskbb} model \citep{mi+84,ma+86}.  
The complete spectral model used was {\sc constant*wabs*(diskbb+zpow)}. The power-law photon index ($\Gamma$) was 
1.55$\pm$0.01 indicating a hard power-law. The inner disc temperature was found to be 155$\pm$3 eV. The spectral fit was reasonably 
good with $\chi^2/dof\;=\;997/768$ and the fitted spectrum with its residual is shown in Figure \ref{fig3}.
However, the inner disc temperature is not redshift corrected. The redshift correction gives a temperature of
246 eV for the redshift 0.5846. 

\noindent Even though the model MCD~+~PL gave reasonably good fit to the observed spectra, the model has a caveat 
that needs to be discussed in the present context. The inner disc temperature obtained here by using the multicolour disc 
blackbody spectrum is consistent with the values reported for other AGN in the literature \citep{wf93, gd04, por+04, cz+03, cr+06, dew+07, min+09}. 
The earlier reported values fall in the range of 0.1--0.2 keV. 
But the inner disc temperature obtained here (0.246 keV) is higher than that is expected for a Shakura-Sunyaev~(SS) disc \citep{ss73} around a black hole of 
mass 3 $\times$ 10$^7M_{\odot}$ estimated by \citet{yu+08} for \pmn. Recently, \citet{don+12} introduced the idea of
colour temperature correction to the AGN disc spectrum. It was shown that this correction is substantial for maximum temperature more than 
10$^5$ K. As discussed by \citet{don+12} that even though the colour temperature correction  is substantial, it is
still not sufficient to explain the soft excess and hence the Compton scattering of photons is essential.
\begin{figure*}
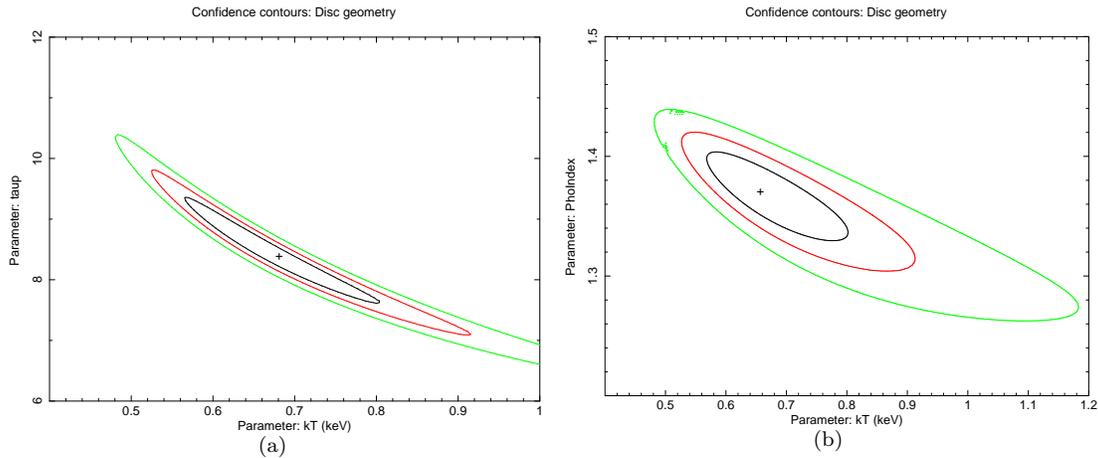

\begin{center}
  \subfigure[]{
    \includegraphics[height=0.3\textheight,angle=-90]{./contour_kT-Tau_disk.ps}} 
  \subfigure[]{
    \includegraphics[height=0.3\textheight,angle=-90]{./contour_kT-phindx_disk.ps}} 
    \caption{68, 90 and 99 percent 2D confidence contours for two pairs of the parameters of 
             {\sc{CompTT}} model with disc geometry; {\it left:} electron temperature and
             opacity and {\it right:} electron temperature and power-law index} \label{fig4}
\end{center}
\end{figure*}

\begin{figure*}
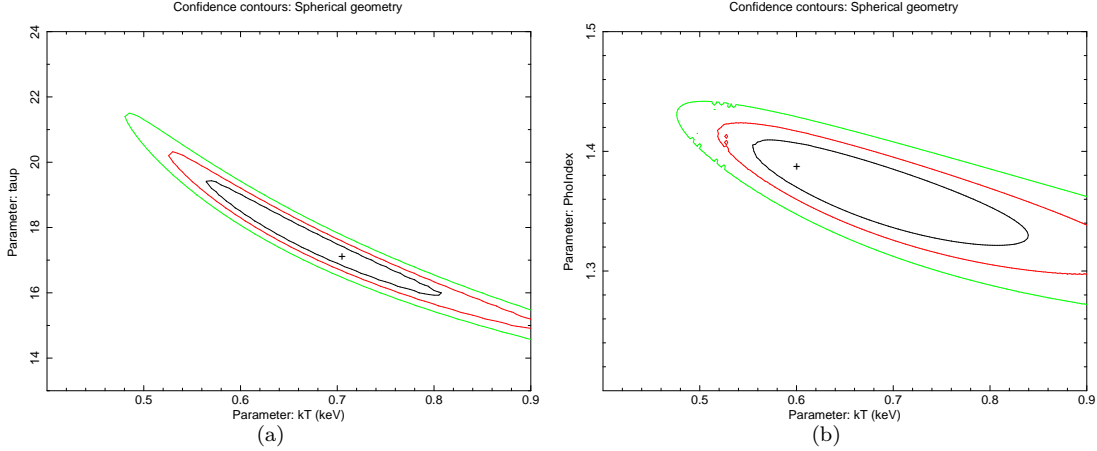

\begin{center}
  \subfigure[]{
    \includegraphics[height=0.3\textheight,angle=-90]{./contour_kT-Tau_sphere.ps}} 
  \subfigure[]{
    \includegraphics[height=0.3\textheight,angle=-90]{./contour_kT-phindx_sphere.ps}} 
    \caption{68, 90 and 99 percent 2D confidence contours for two pairs of the parameters of 
             {\sc{CompTT}} model with spherical geometry; {\it left:} electron temperature and
             opacity and {\it right:} electron temperature and power-law index} \label{fig5}
\end{center}
\end{figure*}

\subsubsection{Thermal Comptonization of disc photons}
\noindent A further possibility can be that the seed photons originating from the accretion disc are Compton boosted
by thermal electrons in an optically thick Comptonizing medium (corona) to produce the soft excess in the observed 
spectrum. The complete model used here is {\sc{constant*wabs*(CompTT+zpowerlaw)}} where the model {\sc{comptt}} 
\citep{tit94} takes care of the thermal Comptonization 
of soft photons and the power-law model is required to fit the hard power-law component in the observed spectrum.
The free parameters in the model are the temperature of the seed photons, the temperature and the optical depth of 
the Comptonizing electrons and the photon index of the power-law component. As the geometry of the emission region 
is not known, we fitted the spectrum for spherical as well as the disc geometry.  The seed photon temperature is kept 
fixed at 30 eV, a value typical for the inner region of the Shakura-Sunyaev accretion disc \citep{ss73} around a black hole of 
mass 3 $\times$ 10$^7M_{\odot}$. For the disc geometry of the Comptonizing plasma, the temperature of the Comptonizing 
electrons came as 0.65$^{+0.15}_{-0.08}$ keV and the photon index of the hard component was 1.37$\pm0.03$ whereas for 
spherical geometry the temperature of the plasma was 0.71$^{+0.10}_{-0.14}$ keV and the photon index was 1.36$^{+0.05}_{-0.03}$. 
Even though the photon index for the power-law component is same for both geometries, the emission region is marginally 
hotter for spherical geometry as compared to the disc geometry. The optical depth for disc and spherical geometries 
are 8.59$^{+0.68}_{-0.86}$ and 17.12$^{+2.23}_{-1.08}$, respectively. For the disk and sperical geometries of Comptonizing
corona, $\chi^2/dof$ came out to be~$932/767$.
The temperature of the Comptonizing plasma came 
out to be higher than that was obtained by using the phenomenological models. Also the photon index of the power-law 
component is harder than that obtained by using phenomenological models. The fitted spectrum for the disc geometry is 
shown in Figure \ref{fig3}. To rule out the possibility that the parameters determined here due to some local minimum 
of the $\chi^2$ function, we used the {\sc Steppar} utility in {\sc Xspec} and explored the parameter space. 
The equal $\chi^2$ contour diagrams for parameter combinations plasma temperature - optical depth and plasma 
temperature - photon index are shown in Figure \ref{fig4} and \ref{fig5}, respectively. This is done for both disc 
and spherical geometry of the Comptonizing plasma. From these figures, it is clear that the values of the plasma temperature 
and  photon index, as obtained by using phenomenological models, can not give acceptable $\chi^2$ values when we 
use {\sc compTT} and {\sc power-law}.  This is true for both spherical as well as disc geometries. In fact, the slab and 
the sphere geometries for the Comptonizing region can not be distinguished statistically.

\subsubsection{Smeared wind absorption model}
Here and in the following subsection we discuss the alternative models which consider an atomic 
origin of the soft excess. \citet{gd04} discussed the relativistic
smearing of characteristic atomic absorption features in a moderately relativistic wind/outflow from the 
inner accretion disc to explain the soft excess in the X-ray spectrum of Seyfert~1 galaxies.
This model is described by {\sc Swind1} \citep{gd04} within {\sc Xspec}. The main parameters of the model are the 
absorption column density, ionization parameter ($\xi = \frac{4\pi F}{n}$, F is the flux of illuminating radiation
and n is the density of the absorbing wind) and velocity dispersion ($\sigma$) of the Gaussian velocity distribution 
of the wind.  The model  {\sc Swind1} along with 
the photo-electric absorption {\sc Wabs} and the power-law for the background continuum gives 
$\chi^2/dof \,=\, 947.32/767$ with the present data from PN and MOS detectors. The fitted parameters are - the 
absorption column density: ($34\pm3) \;\times$ 10$^{22}$ cm$^{-2}$, the ionization parameter : $3.34\pm0.03\;$  erg~cm~s$^{-1}$
and the velocity dispersion $\sigma \,>\, 0.5$ (in the unit of $\frac{v}{c}$) and the photon index of 1.82$\pm$0.01. 

\noindent Even if the fit is reasonably good, the main problem with this model arises from the lower limit of the velocity dispersion 
as obtained from the fitting. The velocity dispersion of the wind, $\sigma\,>\,0.5$ (in terms of $\frac{v}{c}$) 
implied that moderately relativistic wind (with Lorentz factor higher than 1.2) was required to generate the 
necessary smearing of the absorption features so that it could generate the smooth soft excess in the spectrum. 
Similar to the results obtained here, this model typically requires large velocity dispersions when applied to
other sources \citep{gd04}. However, \citet{sd07, sd08, sd09} showed through 
self consistent simulations that it is difficult to generate such winds from accretion disc driven by radiation 
or thermal pressure. The simulations performed by those authors showed that the generated wind can not produce 
sufficient blurring of the absorption features required to generate the shape of the soft excess. 
The remaining alternative is magnetically driven outflow. The magnetohydrodynamic simulations show that a matter
dominated centrifugally driven jet with larger opening angle can be self-consistently produced from the
accretion disc \citep{hw06}. Such jets have low velocity ($\sim$ 0.4c) and also have low density ($\sim$ 500 cm$^{-3}$).
Due to the low density sufficient absorption of radiation is not possible while the low velocity will not provide 
the required smearing of absorption lines to generate the shape of the smooth soft excess in the X-ray spectrum (see
also \citet{sd08}).  
\begin{table}
\centering
\begin{tabular}{llllll}
\hline 
\hline
\multicolumn{1}{c}{Parameters}    & \multicolumn{5}{c}{Best-fit values} \\
\hline
\hline
                      Model      & \multicolumn{5}{l}{constant*wabs*(diskbb + zpowerlaw)}         \\
                 $T_{in}$ (keV)  & $0.155_{-0.004}^{+0.003}$   \\                  
                 $N_{diskbb}$    & $87.36_{-7.22}^{+10.01}$    \\        
                 $\Gamma$        & $1.55\pm0.01$  \\                      
                 $N_{pl}$ ($\rm{photons~keV^{-1}~cm^{-2}~s^{-1}~at~ 1keV}$)       & $1.00_{-0.01}^{+0.02}\times10^{-3}$\\
                 CF              & $1.040\pm0.006$        \\
               $\chi^2 / \nu$    & 996/768  \\
                 log(unabsorbed Flux) (0.3--10 keV)  ($\rm{erg~cm^{-2}~s^{-1}}$)  & -11.355$\pm$0.002 \\
\hline
                   Model          & \multicolumn{5}{l}{constant*wabs*(compTT + zpowerlaw) with spherical geometry}\\
                  $T_0$ (keV) 	 & $3.0\times10^{-2}$ \\
		 $kT_e$ (keV)	 & $0.71_{-0.14}^{+0.11}$\\
		$\tau_{p}$       & $17.12_{-1.08}^{+2.23}$\\
		 $N_{compTT}$ 	 & $5.486_{-0.007}^{+0.011}\times10^{-2}$\\
                 $\Gamma$        & $1.36_{-0.03}^{+0.05}$  \\
		 $N_{pl}$ ($\rm{photons~keV^{-1}~cm^{-2}~s^{-1}~at~ 1keV}$)         & $6.95_{-0.46}^{+0.68}\times10^{-4}$ \\
                 CF              & $1.040\pm0.006$        \\
               $\chi^2 / \nu$    &  932/767 \\
                 log(unabsorbed Flux) (0.3--10 keV)  ($\rm{erg~cm^{-2}~s^{-1}}$)  & -11.339$\pm$0.002 \\
\hline
                   Model          & \multicolumn{5}{l}{constant*wabs*(compTT + zpowerlaw) with disk geometry}\\
                  $T_0$ (keV) 	 & $3.0\times10^{-2}$ \\
		 $kT_e$ (keV)	 & $0.65_{-0.08}^{+0.15}$\\
		$\tau_{p}$       & $8.60_{-0.86}^{+0.68}$\\
		 $N_{compTT}$ 	 & $5.89_{-0.99}^{+0.72}\times10^{-2}$\\
		 $\Gamma$        & $1.37\pm0.03$  \\
		 $N_{pl}$ ($\rm{photons~keV^{-1}~cm^{-2}~s^{-1}~at~ 1keV}$)       & $7.17_{-0.63}^{+0.45}\times10^{-4}$ \\
                 CF              & $1.040_{-0.006}^{+0.006}$        \\
               $\chi^2 / \nu$    &  932/767 \\
                 log(unabsorbed Flux) (0.3--10 keV)  ($\rm{erg~cm^{-2}~s^{-1}}$)  & -11.340$\pm$0.001 \\

\hline
                   Model         & \multicolumn{5}{l}{constant*wabs*swind1*zpowerlaw} \\
		 col. dens. ($\times 10^{22}$ cm$^{-2}$)	 & $34\pm3$ \\
		 $\log\xi$                                      & $3.34^{+0.03}_{-0.04}$  \\
		 $\sigma$ (in units of $v/c$)               	 & $>0.5$ \\
		 $\Gamma$                                  & $1.819_{+0.009}^{-0.006}$ \\ 
		 $N_{pl}$ ($\rm{photons~keV^{-1}~cm^{-2}~s^{-1}~at~ 1keV}$)	                                 & $2.26\pm0.02\times10^{-3}$ \\ 
                 CF                                              & $1.039\pm0.006$        \\
                $\chi^2 / \nu$                                    &  947/767 \\
                 log(unabsorbed Flux) (0.3--10 keV)  ($\rm{erg~cm^{-2}~s^{-1}}$)  & -11.352$\pm$0.001 \\
              
\hline        
                    Model         & \multicolumn{5}{l}{constant*wabs*kdblur(powerlaw+atable(reflionx.mod)}\\
                     Index                                    &    2.99$^{+0.24}_{-0.34}$       \\
                     R$_{in}$ ($\frac{GM}{c^2}$)              &    $<2.06$  \\   
                     Inclination (deg)                        &    $22$ \\     
                     $\Gamma$ (= photon index)                &    1.10$^{+0.05}_{-0.03}$  \\   
                     norm   ($\rm{photons~cm^{-2}~s^{-1}}$)   &    3.39$^{+0.20}_{-0.27} \times10^{-8}$ \\ 
                     Fe/solar                                 &    0.79$^{+0.06}_{-0.07}$\\     
                     Xi   ($\rm{erg~cm~s^{-1}}$)              &    $1171^{+46}_{-65}$ \\     
                     $N_{pl}$ ($\rm{photons~keV^{-1}~cm^{-2}~s^{-1}~at~ 1keV}$)  &    $3.02^{+0.25}_{-0.20} \times10^{-4}$  \\
                     CF                                       &    $1.040\pm0.006$        \\
                     $\chi^2 / \nu$                           &    960/765\\
                 log(unabsorbed Flux) (0.3--10 keV)  ($\rm{erg~cm^{-2}~s^{-1}}$)  & -11.339$\pm$0.001 \\
\hline        
	\end{tabular}
	\caption{The best-fit values of the spectral parameters derived from the spectral fitting of \emph{XMM-Newton} data from OBS B.}
	\label{tab3}
\end{table}

\subsubsection{Relativistically blurred reflection model}
\noindent 
The relativistically blurred reflection model has been used and discussed in the literature to explain the soft excess
in NLS1 galaxies \citep{cr+06, na+11, wa+13}. This model deals with the atomic origin of 
soft excess and does not require the existence of a relativistic wind.
In this model, the cold optically thick accretion disc is illuminated by a power-law distributed radiation and the accretion
disc medium produces Compton scattered reflected photons along with fluorescence lines. 
The X-ray emission then
includes both the illuminating power law and the ionised disc reflection. The fluorescent lines are broadened and 
blurred due to the high velocities of infalling matter in the inner part of the disc as well as 
the strong gravity of the accreting black hole.  

\noindent 
The reflection component of the radiation in the relativistically blurred reflection model was  
introduced by \citet{rf93} and it was further updated by including
more ionization states and more atomic data by \citet{rf05}. In {\sc xspec}, this component is 
implemented as a tabular model, {\sc reflionx}.
The resultant reflected emission is convolved with a \citet{la91} diskline
profile, which accounts for the blurring of emission lines from an accretion disc around a maximally rotating black hole. 
Relativistically broadened iron emission lines had been observed in several 
AGN \citep{tan+95, Nan+07, ris+13} and the line profiles were successfully modeled in those sources. 
In this model, the soft excess is a series of similarly broadened emission lines from lighter elements,
that blend together into a smooth emission feature.
The Laor profile estimation considers that the accretion disc is geometrically thin, flat and has well defined 
inner and outer radii. The radial dependence of disc emissivity is assumed to be a power-law.  
The final model used for the fitting was {\sc constant*wabs*kdblur(powerlaw + atable\{reflionx.mod\})}.
The parameters of the model are inner and outer radii of the accretion disc, iron abundance of the accreting matter, 
spectral index of illuminating power-law radiation, emissivity index 
of the disc, ionization parameter and the inclination of the disc. 
The inclination of the disc is not known for this source. The radio observation of the 
source gives an upper limit for the jet inclination angle to 22$^o$ \citep{doi+06}. Indeed the jet inclination 
angle is not necessarily be the inclination 
of the disc because jet structure can be bent and can have an apparent inclination angle. Due to the lack of 
precise information on inclination angle we kept it
fixed at 22$^o$. This value is reasonable because the Seyfert~1 galaxies generally have low inclination angle.  
The photon index of the power-law component and the illuminating radiation were tied together during
the fitting process. 
The inner radius of the disc was kept free in the fitting process while the outer radius of the accretion disc is 
kept fixed at 400 (in terms of the gravitational radius $\frac{GM}{c^2}$) which is the maximum value accepted by 
the {\sc kdblur} model. In the fitting process, we extended the energy range beyond the observed bandpass of the \xmm.
This was necessary while using convolution model like {\sc kdblur}. The model was evaluated in the energy range 10~eV--500~keV,
whereas the fitting was carried out in the observed bandpass. 
The index for the radial profile of emissivity is fitted to 2.99$^{+0.24}_{-0.34}$. We obtained the photon index of the 
power-law component as 1.10$^{+0.05}_{-0.03}$ and the upper limit of the inner radius of the accretion disc 
was obtained at $2.06$ (in terms of the gravitational radius $\frac{GM}{c^2}$).  
The {\sc reflionx} component had disc iron abundance $0.79^{+0.06}_{-0.07}$ (in solar units),
ionization parameter $1171^{+46}_{-65}$ erg cm s$^{-1}$. The $\chi^2$/dof came out to be 
960/765.  The fitted spectrum along with the residual 
is given in Figure \ref{fig3} and the fitted parameters are given in Table \ref{tab3}. 

\begin{figure*}
\begin{center}
\subfigure[]{
    \includegraphics[height=0.3\textheight,angle=-90]{./lc-hr.ps}} 
  \subfigure[]{
    \includegraphics[height=0.3\textheight,angle=-90]{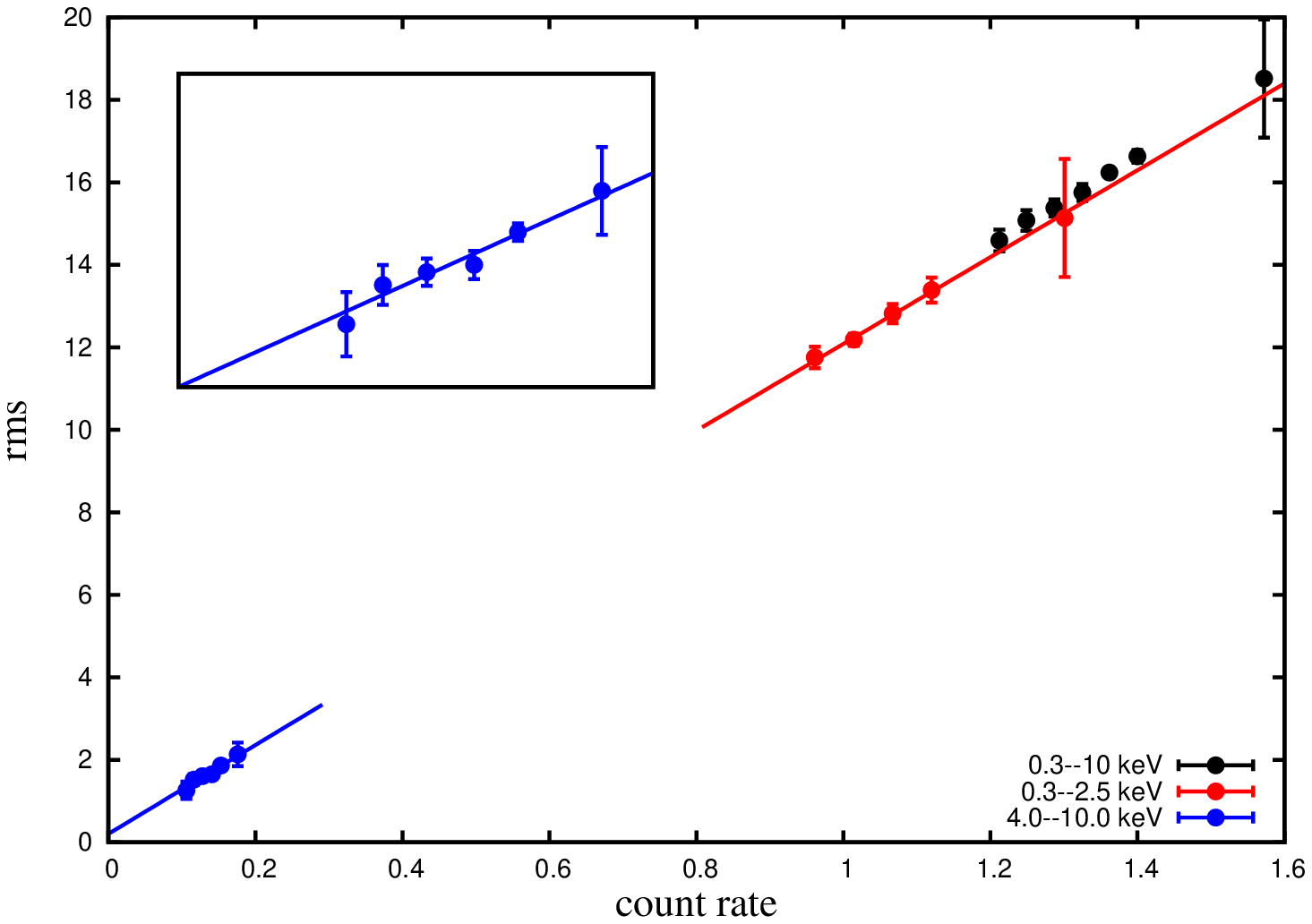}}
    \caption{(a) Lightcurve for 0.3--2.5 keV (S; \emph{upper panel}), 4.0--10 keV (H; \emph{middle panel}) and the corresponding hardness ratio (H/S; \emph{lower panel}) obtained from EPIC-PN data of Obs-B; 
             (b) rms--count rate correlation for \xmm data in the frequency range, $5\times10^{-4}$ -- $5\times10^{-2}$ Hz ; \emph{red}: 0.3--2.5 keV, \emph{blue}: 4--10 keV 
             {\sc powerlaw} component, \emph{black}: 0.3--10.0 keV. Same coloured lines are the best-fit functions for the respective 
	     data sets. \emph{inset}: zoom-in of the {\sc powerlaw} component.} 
    \label{fig7}
\end{center}
\end{figure*}

\subsection{Spectral variability}
The lightcurves with 500 s time bin, in the 0.3--2.5 keV and 4.0--10.0 keV energy bands, and the corresponding hardness 
ratio are shown in Figure \ref{fig7}a. We used 100~s binned lightcurves for the energy ranges 0.3--2.5 keV, 4.0--10.0 keV and for the complete 
energy range 0.3--10.0 keV to calculate the rms amplitude of variability ($\sigma_{rms}$). The rms amplitude of 
variability for a lightcurve with N data points and mean count rate $\bar x$ is given by \citep{umv05}
\begin{equation}
\sigma_{rms} = \sqrt{\frac{1}{N-1}\sum_{i=1}^{N}(x_i\,-\,\bar x)^2}.
\end{equation}
The lightcurve is divided into 2000~s long segments and the $\sigma_{rms}$ is calculated for each segment. The 
variation of the calculated rms variability with the mean count rate for the all three energy ranges are shown 
Figure \ref{fig7}b. It is evident that, within the limited dynamical range of the count rate, the flux vs 
$\sigma_{rms}$ follows a linear relationship. This is true for the lightcurves in all three energy segments. 
The distribution of points are fitted with linear functions with slopes 10.53$\pm$0.58 (for 0.3--2.5 keV band) and 
10.76$\pm$1.41 (for 4--10 keV band).

\begin{figure*}
\begin{center}
  \centering
   \includegraphics[width=0.5\textwidth,angle=-90]{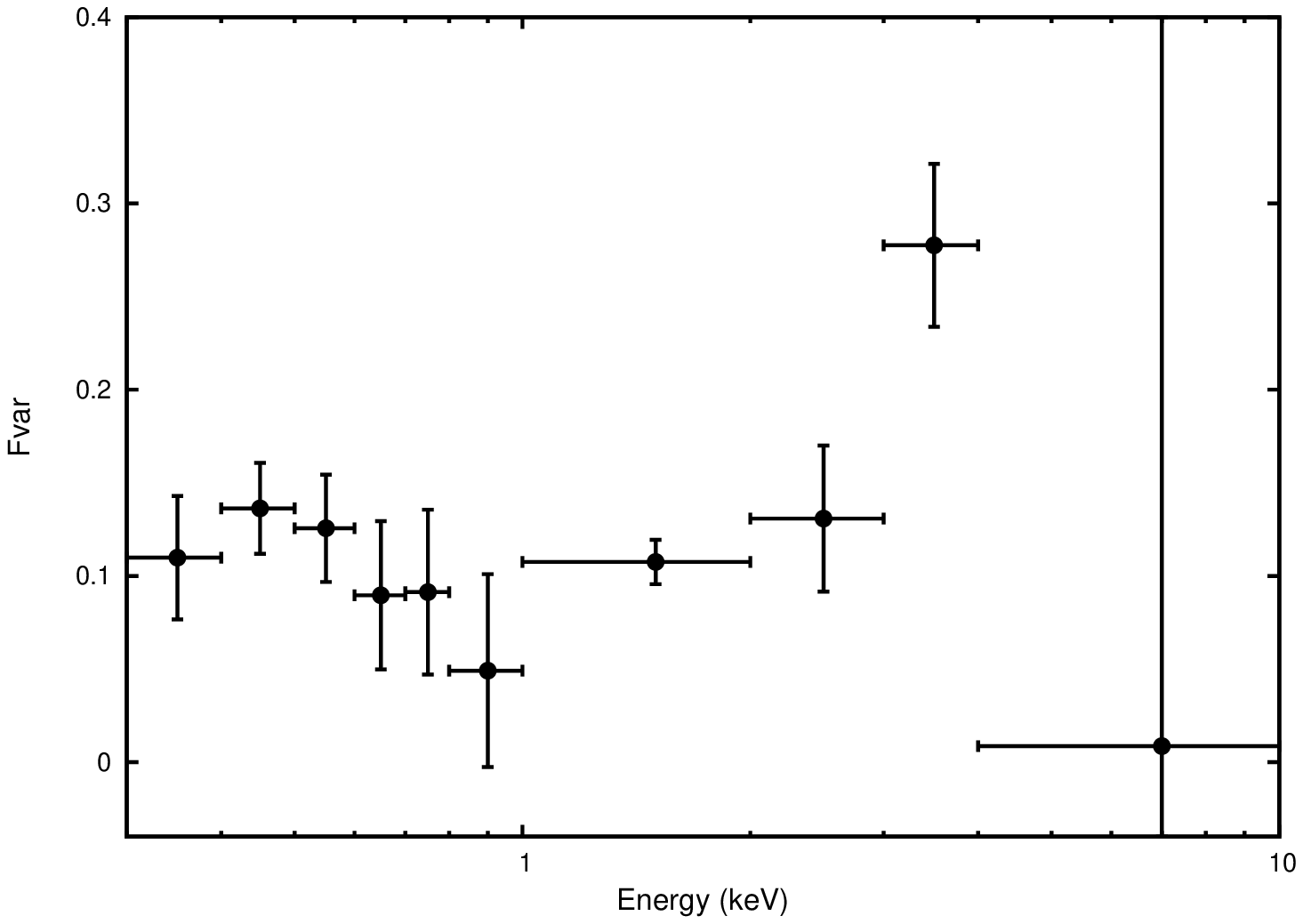} 
    \caption{rms spectrum as obtained from EPIC-PN data of Obs-B.}  \label{fig8}
\end{center}
\end{figure*}

Following \citet{ed+02} and \citet{va+03a}, we calculated the fractional 
variability ($F_{var}$) for the lightcurves in the energy bands 0.3--0.8 keV at the step of 0.1 keV, 0.8--1.0, 
1.0--2.0, 2.0--3.0, 3.0--4.0 and 4.0--10.0 keV. The rms spectrum is shown in Figure \ref{fig8}. The fractional 
variability did not vary significantly up to 1 keV and remained around 10\%. Beyond 1 keV, $F_{var}$ showed a trend of 
increase and reached up to 30\%.

\section{Discussion}
\subsection{X-ray soft excess}
To understand the spectral behaviour of radio-loud narrow line Seyfert~1 galaxy \pmn in the X-ray energy band 
0.3--10.0 keV, we analysed the archival data from \xmm. The EPIC-PN and MOS spectra of the source in 0.3--10.0 keV 
region could not be fitted with a simple power-law. The spectrum in the 2.5--10.0 keV energy range fits reasonably 
well with a power-law distribution with photon index of 1.38$\pm$0.05 while the data in the 0.3--2.5 keV energy 
range show the presence of a distinct soft excess component when compared with the power-law extrapolated from the high energy. 
\begin{figure*}
\begin{center}
  \centering
    \includegraphics[width=0.5\textwidth,angle=-90]{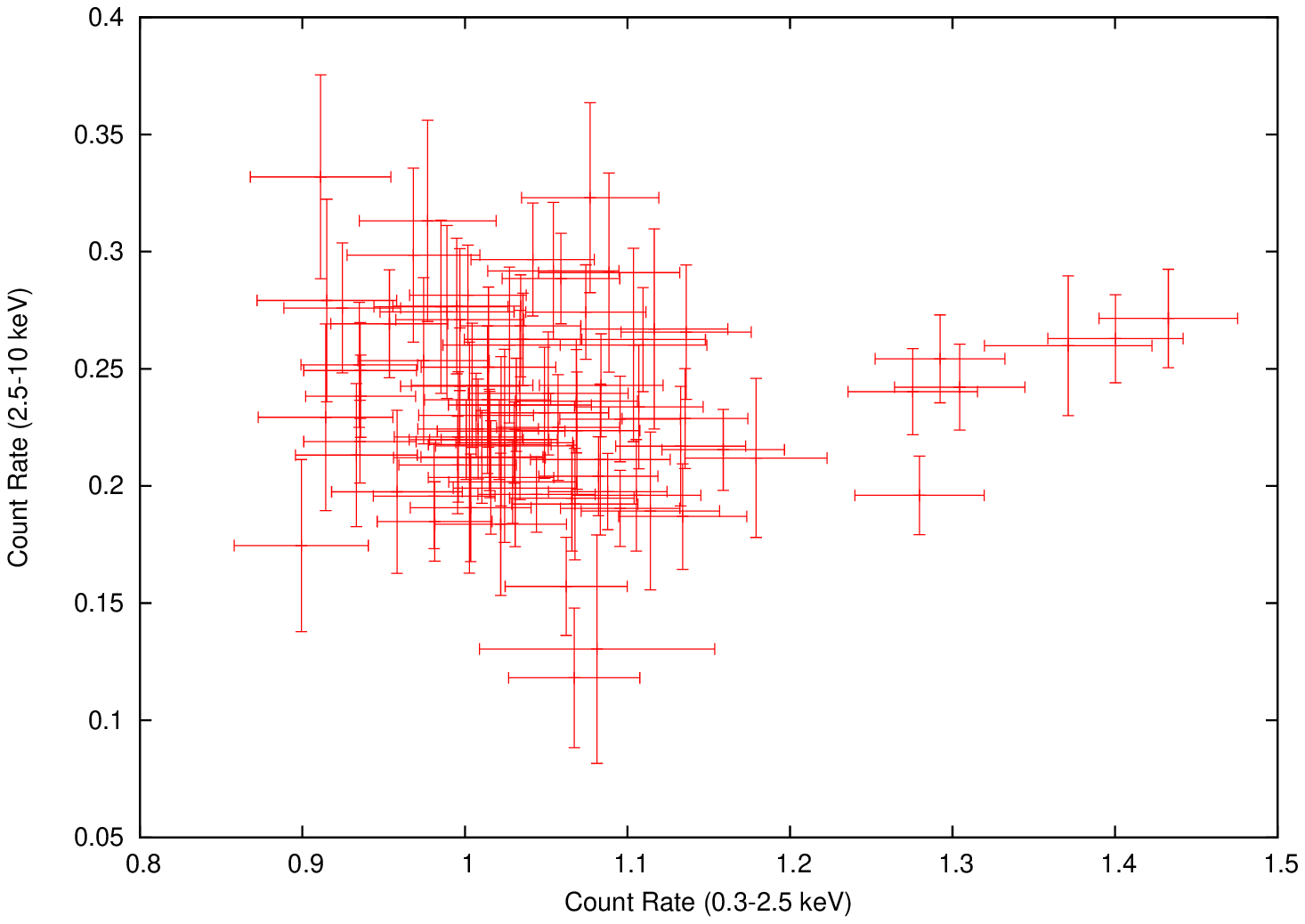} 
    \caption{Flux-flux correlation between 0.3--2.5 keV and 2.5--10 keV lightcurves obtained from EPIC-PN data of Obs-B}\label{fig6}
\end{center}
\end{figure*} 

\noindent Due to the uncertainties in the understanding of the soft excess in the X-ray spectrum of NLS1, 
we considered four different models to fit the EPIC-PN and MOS spectra of \pmn and studied the viability 
of those models. 
Two models, namely direct emission from the disc and optically thick thermal Comptonisation, assume a thermal origin
of the soft excess whereas the other two models, smeared wind absorption and ionised reflection, invoke an atomic origin of 
the soft excess. The multicolor disc blackbody models gave inner disc temperature higher than expected for a Shakura-Sunyave 
disc \citep{ss73} around a black hole of mass $\sim$ 10$^7$M$_\odot$. This makes the direct disc emission model less viable as an 
explanation for the soft excess. In the case of the thermal Comptonization model, the soft excess in the 0.3--2.5 keV energy 
range was modeled by the Comptonization of disc photons in an optically thick cold thermal corona. The 
characteristic photon temperature from the accretion disk was held fixed at 30 eV which is the typical inner disc 
temperature for an accretion disc around a black hole of mass 10$^{7}$M$_{\odot}$. The fitted value of the corona 
temperature was found to be 0.71 keV for the spherical geometry and 0.65 for the disc geometry. The optical depth 
of the corona came as 17.12 for spherical and 8.6 for disc geometries. The fitting of the power-law component
yielded a hard power-law with photon index 1.37 for both the geometries. This hard power-law component can be 
modeled by invoking a hot corona with moderate optical depth ($\tau\,\lesssim\,1$). Such two corona scenarios were 
discussed in the literature in the context of explaining the X-ray spectra of Seyfert~1 galaxies \citep{zd+96, ma+98, dew+07, la+97}. 
The seed photons for the cold optically thick corona are possibly the photons from the accretion disc while the 
seed photons for the hot corona can be either from the accretion disc or the photons of the soft excess from 
the optically thick corona or from both. To distinguish the possible origin of the seed photons for the hot 
corona, we studied the flux-flux correlation between the soft excess and power-law flux. The flux-flux distribution 
does not show any correlation between the two energy bands (Figure \ref{fig6}). If the soft excess photons served as the 
seed photons for the hot corona, there should have been a correlation between these bands. Therefore, it is most 
likely that the disc photons of
temperature 30 eV served as the seed photons for the hot optically thin corona.
As the observed spectrum does not show any cut-off 
within the observed energy band, the temperature of the electrons in the hot corona, which determines
the cut-off in the photon spectrum, can not be constrained. Therefore, the future observations of the source with NuSTAR \citep{har+13} and 
Astro-H\footnote{http://astro-h.isas.jaxa.jp/} in hard X-rays may be required to reveal such a cut-off, if the 
power-law was indeed produced by the thermal Comptonization. High energy observations would also help constrain the
reflection interpretation.   

\noindent As shown by \citet{gd04}, the thermal origin of soft excess leads to a difficulty that the temperatures of the thermal
electron required to fit the spectrum fall within a very narrow range (0.1--0.3 keV) irrespective
of the mass of the accreting black hole. Recently, \citet{jin+12, don+12}
modeled the X-ray spectra of a large number of unobscured type 1 Seyferts using the
thermal Comptonization model and it was found that the electron temperature varies from
0.1 to 0.6 keV, while the black hole mass ranges from $10^6$--$10^9$ M$_\odot$. The values
of electron temperature obtained here are generally consistent with those for other 
NLS1 galaxies reported in the literature.
Such a narrow temperature range and its independence on black hole mass suggest
that the soft excess possibly does not have thermal origin, instead it could be of atomic origin \citep{gd04}. 
We have already discussed previously that the smeared wind absorption model can not account for the
smooth X-ray soft excess as shown by the detailed simulation studies \citep{sd07,sd08,sd09}.
   
\noindent As the last possible alternative we studied the relativistically blurred reflection model which considers
that the soft excess has atomic origin \citep{cr+06, na+11, wa+13}. It considers an accretion disc and a hot corona 
which produces the power-law component by Comptonization
of the soft photons from the disc. The power-law component irradiates the disc producing reflected emission, which
is then blurred due to the strong gravity and high velocities of the accreting material. The fitting process, 
yielded the hard power-law component with photon index at 1.1. Naturally, the illuminating radiation also had 
the same photon index. The upper bound of the inner radius is 2.06 (in terms of gravitational radius
of the central black hole) and the outer radius is fixed at 400 (in terms of gravitational radius
of the central black hole). The index for the radial pofile of emissivity came as 2.99. Generally for its value 
greater than 3.0, it is considered that the emitted radiation is strongly affected by the gravity of the central 
black hole whereas a value equal to 2.0 implies that 
the radiation contributed from the larger disc radii is non-negligible. Thus the present value of 2.99 implies that the 
illuminating corona has a fairly large height above the accretion disc so that the radiation is not affected by the
gravity of the central black hole.  
The corona which resides at a relatively larger height above the accretion disc, may act as the 
base of the jet whose presence is confirmed by the detection of radio and $\gamma$-ray photons from \pmn. 
Such a scenario was discussed by \citet{mark05} and also by \citet{mill12} in the context of galactic black hole X-ray
binaries. 
\citet{wa+13} showed that the radio-loud AGN have weak reflection features in the X-ray spectrum and this is expected if 
the accretion disc corona acts as the base of the jet  \citep{mark05}. 

\noindent Following \citet{wa+13}, we estimated the reflection strength for this source. For this purpose we used the ionised
reflection model {\sc pexriv} where the main parameters are the disc inclination, photon spectral index, iron abundance
(in solar unit), ionization and reflection  parameters. The values of the photon spectral index, the iron abundance and the
ionization parameter in {\sc pexriv} model were freezed to the values obtained with relativistically blurred reflection model. The reflection parameter
was set to zero. The normalization was then adjusted to a value so that the model flux estimated in 2--10 keV energy range 
matches with the observed power-law component flux in the same energy band. Then the reflection parameter was varied
such that the total flux (Compton reflection + power-law) in the 15--50 keV energy range was equal to the flux obtained 
from relativistically blurred reflection model in the same energy band. The final value of the reflection parameter was found to be 0.367.
This is consistent with the fact that the radio-loud AGN have relatively weak reflection component as compared to the radio-quiet
AGN \citep{wa+13}.    

\noindent The corona may well be outflowing as discussed by \citet{belo99} in their dynamical corona model. 
It is shown by \citet{belo99} that a dynamical corona emits a hard power-law, hence the hard power law obtained here 
may be a natural consequence of the dynamical corona at the base of the jet.

\noindent We also studied the rms spectrum from the observed lightcurve. The calculated rms spectrum remained almost constant up to 1 keV
and it increased beyond 1 keV. Thus beyond 1 keV, there might be independently varying spectral components and the power spectral density could
be energy dependent \citep{va+03a}. As the error bars are relatively large,  longer observations are required to draw any firm conclusion on spectral variability for photon 
energies beyond 1 keV. In fact the Fig \ref{fig6} shows no correlation 
between the lightcurves in 0.3--2.5 and 2.5--10.0 keV energy bands, hinting that the lightcurves are possibly of independent origin.  

\noindent Another important aspect of the emission from \pmn is the linear relation between the rms variability amplitude and
the mean count rate. This behaviour is also seen in other NLS1 galaxies \citep{ed+02, mc+04} and from Seyfert~1
galaxies \citep{va+03a, va+03b}. \citet{umv05} showed that the linear rms--flux relation is an outcome of a multiplicative 
process where the amplitude of short time-scale variations are modulated by the longer time scale variations. In the case of an accretion flow, the
long time scale variations are produced at larger radii whereas the short time scale variations are produced at shorter radii of the accretion disc. 
\citet{ly97} proposed a model where the fluctuations originate at the larger radii and then propagate inwards 
to smaller radii to explain the observed X-ray variability in X-ray binaries. \citet{um01} showed that the linear relationship between 
the rms variability amplitude and flux is a natural outcome of the model proposed by \citet{ly97}.    
\begin{figure*}
\begin{center}
  \centering
   \includegraphics[width=0.5\textwidth,height=0.48\textheight,angle=-90]{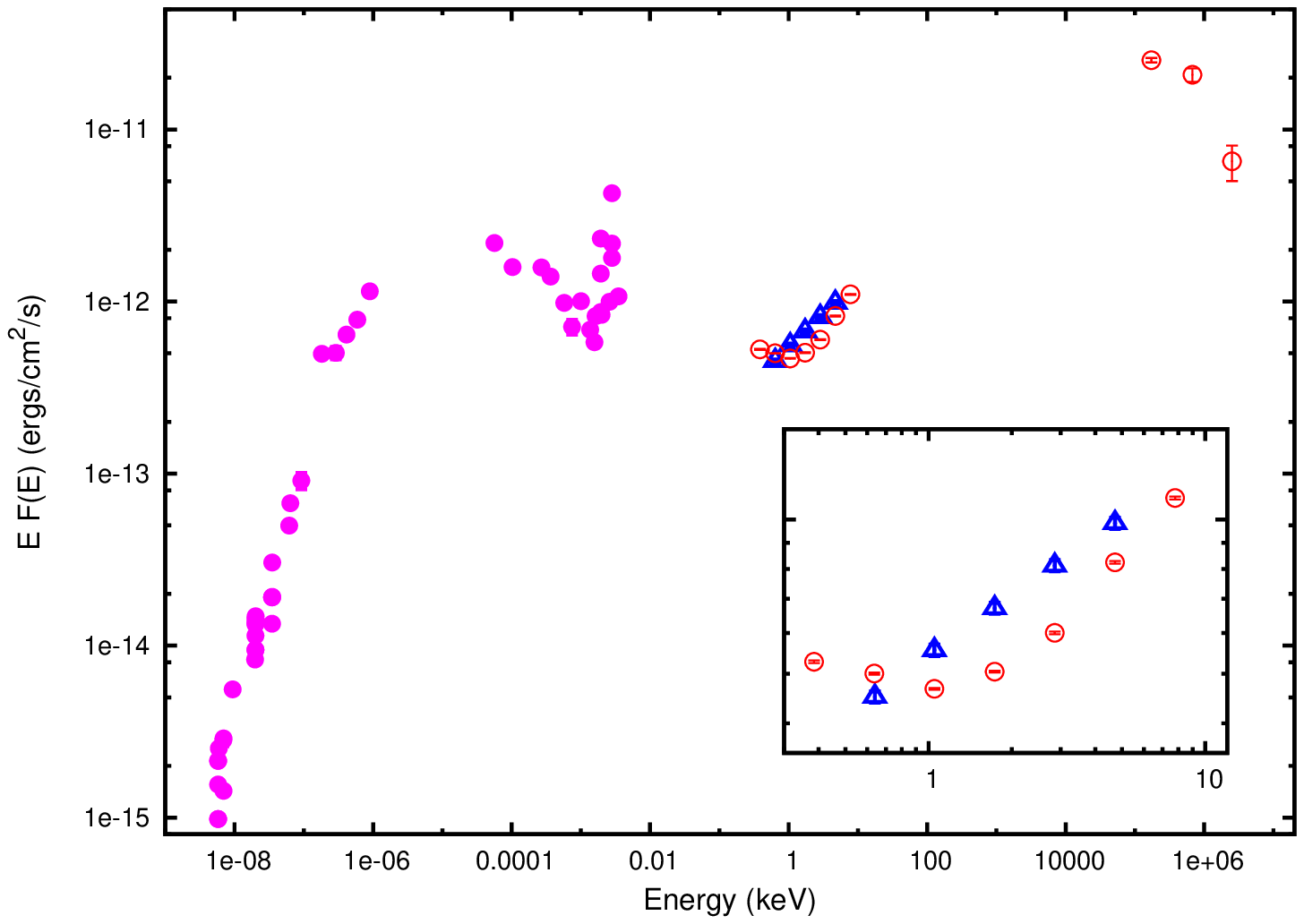} 
    \caption{Broadband spectral energy distribution (SED) of PMN~J0948+0022; \emph{magenta}: archival data taken from ASDC, 
             \emph{red}: spectral points from the present analysis of \emph{XMM-Newton} (Obs-B) and 
             \emph{Fermi}/LAT data, \emph{blue}: spectral points from the present analysis of \emph{Swift}/XRT data.}  \label{fig9}
\end{center}
\end{figure*}
\subsection{Spectral energy distribution (SED)}    
\noindent We analysed the six months \emph{Fermi}/LAT data of \pmn and obtained the flux in the energy
ranges 0.1--0.40 GeV, 0.40--1.5 GeV and 1.5--5.5 GeV. The \emph{Fermi}/LAT data points and the flux values obtained from
\emph{XMM-Newton} PN and MOS detectors in 0.3--10 keV energy band are plotted along with the archival data in the radio-to-optical
band taken from ASDC\footnote{http://www.asdc.asi.it}. The data points from \emph{Swift}/XRT as obtained from the present analysis 
are also plotted for comparison. The effect of
X-ray soft excess is clearly visible in the shape of the SED as shown in Figure \ref{fig9}. The broadband SED of \pmn was previously reported by \citet{abdo+09b}
and \citet{fos+09} using the X-ray data from \emph{Swift}/XRT. The SED in those works were modeled with 
self-synchrotron Compton and external Compton (EC) processes in the X-ray and MeV--GeV band, respectively. But the present SED clearly 
shows that the SED in the X-ray band is not a simple power-law and the presence of the soft excess needs to be accounted for  
to fully understand the overall energetics of the source. The modeling of the SED is beyond the
scope of this work and it will be discussed in a separate work.
     
\section{Conclusions} 
We analysed the archival X-ray data from \xmm, {\it Swift}/XRT and  $\gamma$-ray data from \fermi/LAT for the radio loud 
narrow line Seyfert~1 galaxy PMN~J0948+0022. Our conclusions are as follows. 
\begin{center}
\begin{itemize}
\item The spectral analysis of X-ray data from \xmm establishes the presence of soft excess in the 0.3--2.5 keV energy band
      and a hard power-law beyond 2.5 keV. The near simultaneous {\it Swift}/XRT observation does not reveal this soft excess
      due to poor count statistics.
\item Among four different models used to fit the \xmm spectra of the source, the Comptonization model as well as the relativistically blurred 
      reflection model give physically acceptable description of the spectrum. Both the models assume the presence of a hard power-law emitting
      thermal corona above the accretion disc but its verification needs simultaneous observations of the source in hard X-rays also. 
\item The SED of the source clearly shows the presence of soft excess which was not reported earlier. This component can not be ignored while 
      modelling the SED of the source.  
\end{itemize}
\end{center}

\section*{acknowledgment}
The authors acknowledge the anonymous reviewer for suggestions which improve the quality of the work substantially.
Authors also acknowledge Andy Fabian for providing the extended version of Reflionx model. 
HB acknowledges support from Department of Science \& Technology INDIA, through INSPIRE faculty fellowship IFA-PH-02. 
HB further acknowledges R. C. Rannot and R. Koul for their support to work and to pursue DST-INSPIRE position at 
ApSD, BARC, Mumbai.
This research has made use of data obtained from the High Energy Astrophysics Science Archive Research Center (HEASARC), provided by 
NASA's Goddard Space Flight Center.

\bibliographystyle{mn2e}
\bibliography{ms-v6-05Jan14}

\end{document}